# Epistemic Networks

Mihnea C. Moldoveanu
Rotman School of Management, University of Toronto
Email: mihnea.moldoveanu@rotman.utoronto.ca

Joel A.C. Baum
Rotman School of Management, University of Toronto
Email: joel.baum@rotman.utoronto.ca

**Abstract**

We show how important phenomena in social networks like coordination, trust and the communication of unsubstantiated information ('gossip') – can be modelled and understood using epistemic networks or *epinets:* directed graphs comprising networked agents and the key facts, statements or other kinds of propositional beliefs ('propositions') relevant to their actions. We use *epinets* to sharpen the explanatory and reach of social network analysis to situations problematic to both network-structural approaches and game theory.

The dynamics of human groups are shaped not only by interpersonal ties and the topology of these connections but also by *interactive beliefs* – what each person believes about what every other one believes. Mobilization [Chwe, 1999] and coordination [Schelling, 1978] are enabled by interactive belief structures such as 'I think you will mobilize if I do' or, 'I think you think I will mobilize if you do': they undergird and rationalize collective action. Social network analyses focus on peer-to-peer relationships among agents (eg. interaction, friendship, information flow) and network-specific properties of agents (eg. betweenness, eigen-, closeness centrality) or sub networks (eg gamma-neighborhoods; network constraint, cliqui-ness) but rarely the epistemic and interactive dimensions of social groups (What *kind* of information flows? To whom? Who else knows? Who knows about who else knows?).

Both models of social networks based on node interaction and link creation rules [Laszlo Barabassi and Albert, 2002] and game-theoretic models that feature interactive belief hierarchies have trouble explaining social phenomena wherein (1) interactive belief systems matter, but (2) common knowledge or common belief of agents' rationality, strategies and payoffs is not accessible or easily achievable[1]. By taking a syntactic view of the state space (in terms of *statements about events* rather than about self-evidently interpretable *events*) [Halpern 2015]), epinet representations resolve ambiguities arising from different agents using different language systems to refer to qualia and un-propositionalized perceptions or sense experiences [Shin, 1993]. They capture salient details of situations in which the way one represents an event makes a causal difference to the way in which people respond to it [Halpern, 2015, ibid.]. This feature gives modelers more degrees of representational freedom: the language system used to represent events via propositions is itself a variable to be specified. Ambiguity about ways in which sentences attach to states of affairs still lurks. But it is no longer hidden by a formalism that assumes all agents have 'the same' interpretation of an event – the same representation of raw perceptions as propositions about events or objects.

We introduce *epistemic networks* to model interactive systems of knowledge and belief in social groups, and represent individual and collective states of uncertainty, confidence, oblivion, awareness, mutual knowledge and trust. To see how an epinet is built, consider this example. Alice is seeking a job in Bob's firm. Bob interviews her for this purpose. Alice had made some false statements on the resume now sitting before Bob while the interview is proceeding – and a notable one about having won an engineering contest in her college in the year *2018*. Alice is unaware of the fact that Bob knows that his *own* son had actually won an engineering contest that same year at the same college. Nor does she know – or know she does not know - that Bob believes the contest he knows his son to have won is the *same* contest Alice claims to have won. Bob suspects Alice of lying. Alice believes her deception was successful. As the interview unfolds, Alice thinks Bob believes her. But he does not. He infers from her facial expression she believes her deception was successful and he is right. Alice is confident of the success her deception. She puts any complications out of her mind because she is *oblivious* to Bob's epistemic state. An epinet representation (Figure 1) links Alice and Bob to specific beliefs (p, q) that 'matter to the situation'. Let *q* denote the proposition 'Alice won the engineering contest in year X' and *p* denote its negation. Alice believes *p*, believes Bob believes it, and that he believes she believes it. But Bob knows *q* and believes – through inference - that Alice believes he does not believe it. The epistemic network below represents the $1^{st}$ (Alice knows p, Bob knows p), $2^{nd}$ (Alice believes Bob believes q, Bob believes Alice believes p) and $3^{rd}$ (Alice believes Bob believes she believes q, Bob believes Alice believes he believes p) that are jointly explanatory of the epistemic structure of their interaction.

---
[1] The absence of negative introspection - eg. not knowing specific elements of an interactive state space - even if they are of low probability - poses challenges for partitional state spaces [Dekel, Lippman and Rustichini, 1998] that cannot be adequately resolved by standard game theoretic frameworks: Roughly: if I do not know $p$ and do not know I do not know it and you know $p$, then we cannot equilibrate our belief systems to achieve an equilibrium outcome without (credible) pre-play communication.

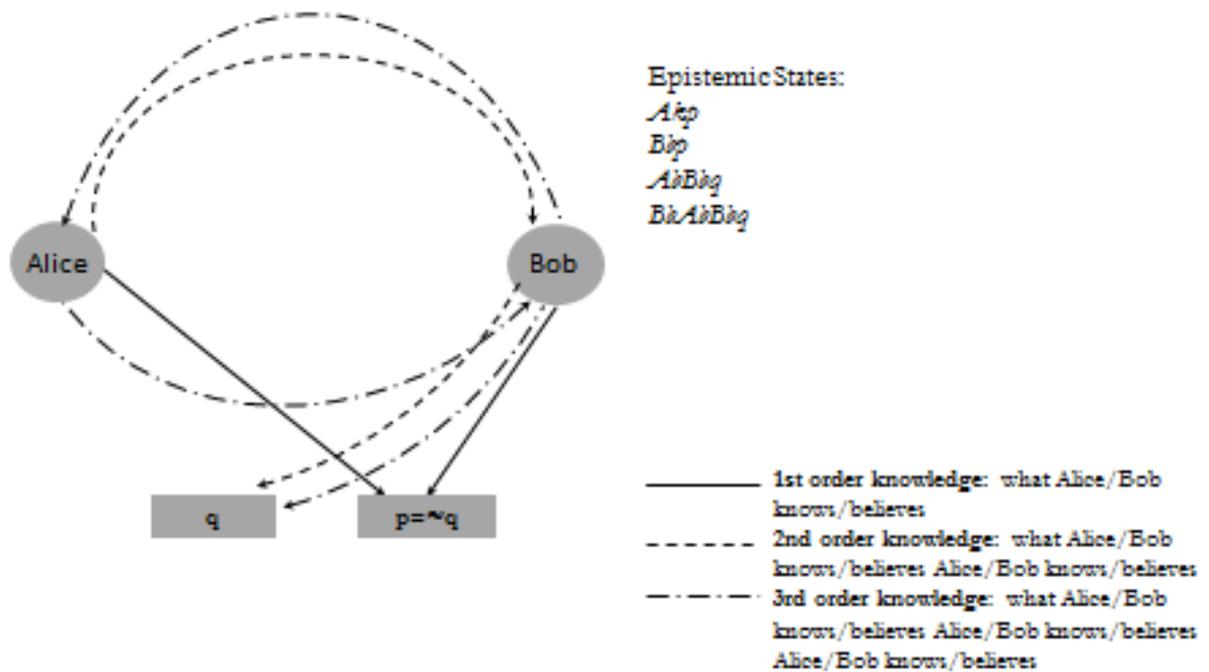

Figure 1. Epistemic Network for Alice and Bob Fake CV Interview Example.

The core building blocks of an epistemic network are not only individual agents and the propositions relevant to their interaction(s) but also their epistemic states *vis a vis* these propositions and each others' beliefs [Figure 2A]:

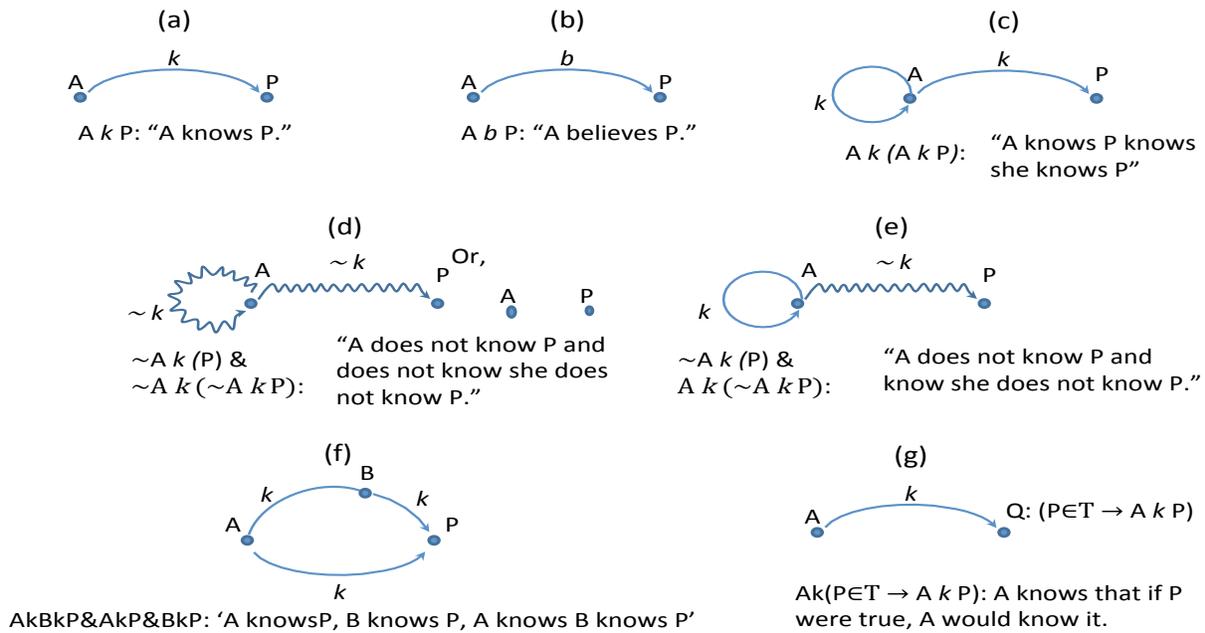

*Figure 2.A. Epistemic State Representations for Dyadic Epinets.*

*Knowledge (K)*: AkP 'A knows P', for some agent A and some proposition P. If agent A knows, for instance, P='a judgment in favor of the defendant was entered today' then agent A will, *ceteris paribus*, act as if P is true in those cases in which (a) P is relevant and (b) A sees P as relevant (in which case we will say that P is salient to A). k is a simple binary relation (either AkP or ~AkP) between an agent and a proposition P that is characterized by the following necessary conditions: (a) A believes P, and (b) P is true[2]. *Belief* is a weaker state: AbP ('A believes P') simply represents the weaker (first prong) of the conjunction above.

*Awareness and Awareness of Level n ($k^n$)*: 'A knows P & knows he knows P' is a state that we will refer to as awareness. 'A knows he knows P, and so forth, to n levels': AkP, Ak(AkP), Ak(Ak(AkP)), Ak(....AkP)...), abbreviated $Ak^nP$ generalizes awareness to a state in which A knows P, knows that she knows it, and so forth. This state relates to an agent's introspective insight into the contents of her knowledge. It is possible that A knows P but does not know he knows P (i.e.,

---

[2] These are not *sufficient* conditions for knowledge [Gettier, 1963]. *A* may believe *P='There is a quarter in my pocket'* on the basis of the vague sensation of a weight in his left pocket, and there is, indeed, a quarter in *A*'s pocket, but it is in his right pocket, not his left one. In this case, *A* does *not* know *P*, even though he has a valid reason for a true belief. What counts is having the *right* reason for holding the belief, i.e. a reason that may be causally related to the proposition expressed by the belief being true.

*AkP&~(Ak(AkP))*. Call this state *unawareness* of P.

*Ignorance*: 'A does not know P but, but knows she does not know P': ~(AkP) & Ak(~AkP). Ignorance of P (which can also understood as uncertainty about P) is such that A pays attention to the values of variables relevant to the truth value of P, but does not know those values or the truth value of P.

*Oblivion*: 'A does not know P, does not know that he does not know it, and so forth': ~(AkP)&~(Ak(~AkP))& …. In this state, A does not know P nor is *heedful*, in any way, of P, or of information that could inform A about the truth value of P. She does not pay attention to variables or experiences that could impact the truth value of P. Whereas an ignorant A will raise questions about P when P is relevant, and an A who is level-2-aware of his ignorance of P will raise questions about P when P is relevant, an A who is oblivious of P will do neither.

Our definition of a state of knowledge can be augmented by an *epistemic qualifier* that safeguards an agent's epistemic state and functions as a warrant for action predicated on that state. An agent may believe that a proposition P is true to a certain degree w, but, within the set of beliefs that an agent attaches a similar degree of credence to, there is an important distinction to be made between *w*-strong belief and belief that *were P true* (or, false), then the agent would know or come to know it. This *subjunctive* epistemic state relates to the confidence that the agent has in his or her own beliefs. Instead of modeling this state as a second-order probability or degree of credence (a degree of credence in the degree with which the agent lends credence to P), we model it via:

*Confidence* (Con): 'A knows that if P were (not) true, A would (not) know P': Ak( $"P \in T \rightarrow AkP"$ ), where the double quotation marks are meant to signify the modal form of the enclosed phrase. Confidence is a binary measure - which can be turned into a continuous measure if necessary - that can be used to answer the question: *Does A believe what she knows*? The converse question – i.e. *Does A know what she believes?* - is captured by A's awareness.

Social network-relevant social situations feature *collective* epistemic states relevant to the propensity of individual actors to co-mobilize ('I will go if (I know that) you will go (if I go)') and co-ordinate ('I will come to meet you at place X at time t if I think you will too and I think you think I will too'). Epinets represent collective states as follows [Figure 2B]:

*Distribution or Shared-ness*: 'A knows P and B knows and C knows P and ….): AkP & BkP & CkP & DkP &… Distribution measures the spread of knowledge of P in a network G and can be measured in absolute (the total number of agents in G that know P) or relative terms (the proportion of agents in G that know P).

*Collective Awareness of level n: A knows$^n$ P& B knows$^n$ P & C knows$^n$ P&…* Collective awareness measures the degree of level-n awareness about P in the network G, either as the number of agents that *know$^n$ P* or as the proportion of agents in G that *know$^n$ P*. Second-level collective awareness ($k^2$) is a condition that guarantees agents in G that know P will find P salient when P is relevant (because, they *know* they know P). Awareness guarantees that the agents in a network know *P* and *find P salient* to an interaction *when P is relevant* to a problem or issue they are interested in.

*Near-Commonality of Level n* ($NC^n$): '*A* knows *P*, *B* knows *P*, *A* knows *B* knows *P*, *B* knows *A* knows *P*, and so forth, to level n': (*AkP*) & (*BkP*) & (*AkBkP*) & (*BkAkP*) &..., abbreviated: $(AkBk)^n(P)\&(BkAk)^n(P)$. Commonality of level *n* measures the level of interactive knowledge of *P* of actors in *G* and is a measure of the *coordinative potential* of the network. Many coordination scenarios require not only actor-level knowledge of focal points or coordinative equilibria and a selection criterion among them, but also actor-level knowledge about the knowledge of other actors and about the knowledge that other actors have of the actor's own knowledge [Schelling, 1978]. *Level-2* commonality (mutual knowledge), and level-3 commonality (*AkBkAk(P)&BkAkBk(P)*) of knowledge about some proposition *P* are particularly important for studying network-level mobilization and coordination.

Social networks are clique-y [Wasserman and Faust, 1994; Burt, 1992]: they contain sub-networks that are connected by ties that encode friendship, trust, influence and information flows – or even the probability of emotional contagion. We *assume* a clique has a better chance of coordinating because it possesses a higher level of 'cohesion, but 'cohesion' is not something we directly measure. Rather, we *infer* it from connectedness. Let us take 'cohesion' to mean 'epistemic coherence of first (what I know or think to be true and what you know or think to be true), second (what I think you think regarding the truth value of some proposition P) and third order beliefs (what I think you think I think about truth value of that proposition). Take some proposition *P* whose truth value is independently known, Alice and Bob form a cohesive dyad just in case Alice knows *P* (which means that *P* is true by the definition of the *knows* operator) knows that Bob knows *P*, and knows that Bob knows she knows *P*, while, at the same time, *B* knows *P*, knows that Alice knows *P*, and knows that Alice knows he knows *P*. Cohesiveness can be a matter of degree: Alice may know *P* and know that Bob knows it, but may not know that Bob knows she knows it – and vice-versa. Given this epistemic gap allows Alice and Bob may be able to act cohesively in some situations but not in others - a difference in individual states that makes a difference to joint action.

To test the representational power of epinets, we studied the correlations between network structure, beliefs about network structure and interactive beliefs about a relevant set of propositions in a network of academics trained in game theory. We constructed a questionnaire (**Supplemental Information: 1**) to measure both 'the network' and the associated epinets around 'issues' - represented by positions with well-defined truth values - relevant to coordination and co-mobilization of the agents. From answers received, we constructed the *collaboration, friendship and interaction* networks of faculty members in a disciplinary area of a North American graduate school of management, and measured the betweenness, degree and eigenvector centrality of each faculty in each of the networks (Figure 3.A-3C).

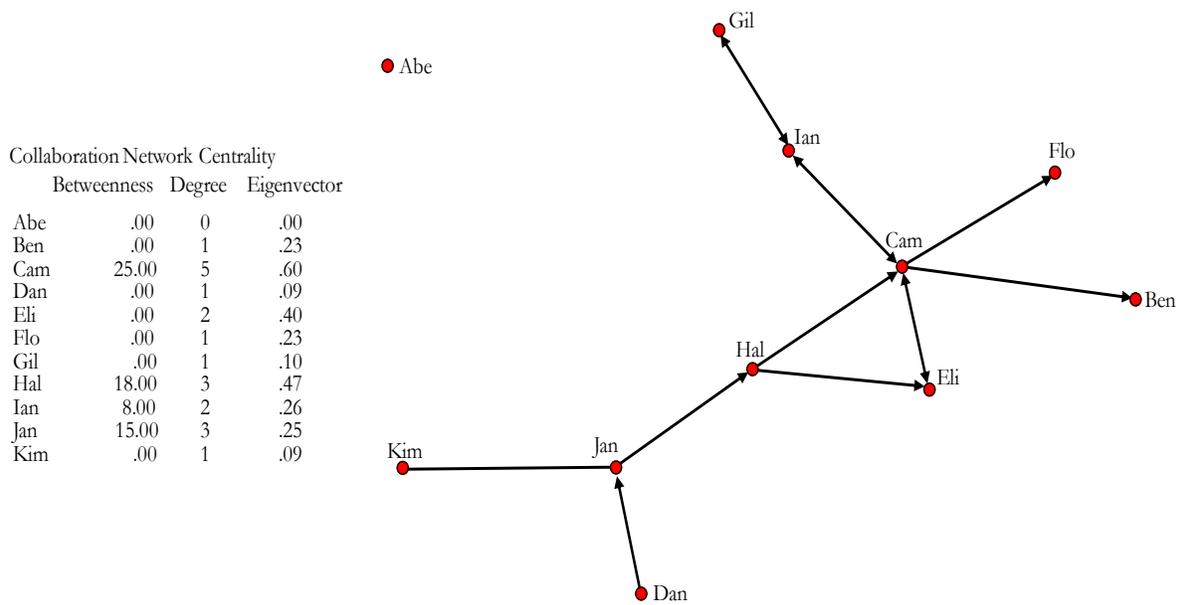

| Collaboration Network Centrality | | | |
|---|---|---|---|
| | Betweenness | Degree | Eigenvector |
| Abe | .00 | 0 | .00 |
| Ben | .00 | 1 | .23 |
| Cam | 25.00 | 5 | .60 |
| Dan | .00 | 1 | .09 |
| Eli | .00 | 2 | .40 |
| Flo | .00 | 1 | .23 |
| Gil | .00 | 1 | .10 |
| Hal | 18.00 | 3 | .47 |
| Ian | 8.00 | 2 | .26 |
| Jan | 15.00 | 3 | .25 |
| Kim | .00 | 1 | .09 |

*Figure 3A. Centrality Measures for Collaboration Network.*

**Friendship Network Centrality.**

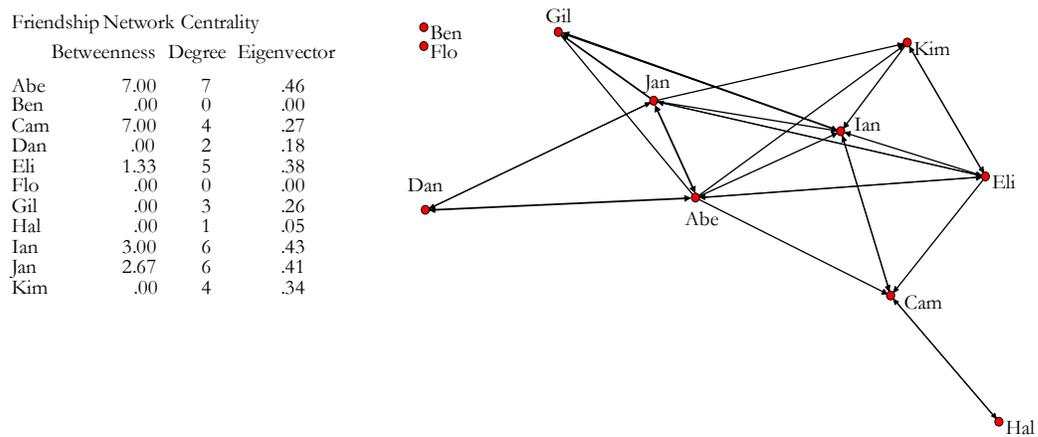

| Friendship Network Centrality | | | |
|---|---|---|---|
| | Betweenness | Degree | Eigenvector |
| Abe | 7.00 | 7 | .46 |
| Ben | .00 | 0 | .00 |
| Cam | 7.00 | 4 | .27 |
| Dan | .00 | 2 | .18 |
| Eli | 1.33 | 5 | .38 |
| Flo | .00 | 0 | .00 |
| Gil | .00 | 3 | .26 |
| Hal | .00 | 1 | .05 |
| Ian | 3.00 | 6 | .43 |
| Jan | 2.67 | 6 | .41 |
| Kim | .00 | 4 | .34 |

*Figure 3B. Centrality Measures for Friendship Network.*

## Interaction Network Centrality.

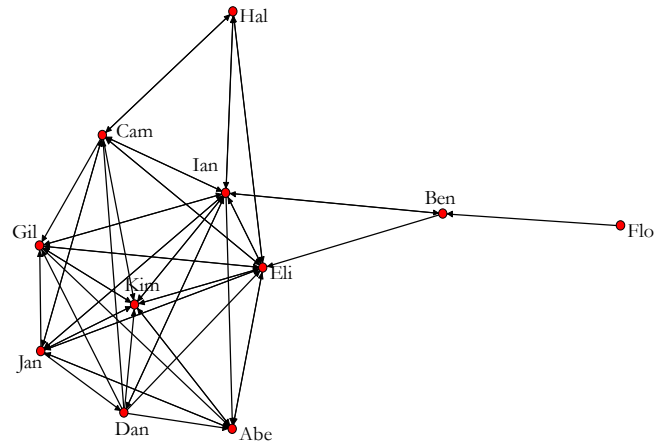

Interaction Network Centrality

|     | Betweenness | Degree | Eigenvector |
| --- | ---: | ---: | ---: |
| Abe | .00 | 6 | .31 |
| Ben | 9.00 | 3 | .11 |
| Cam | 1.33 | 7 | .33 |
| Dan | .17 | 7 | .35 |
| Eli | 9.00 | 9 | .38 |
| Flo | .00 | 1 | .02 |
| Gil | .17 | 7 | .35 |
| Hal | .00 | 3 | .15 |
| Ian | 9.00 | 9 | .38 |
| Jan | .17 | 7 | .35 |
| Kim | .17 | 7 | .35 |

*Figure 3C. Centrality Measures for Interaction Network.*

To this traditional network picture, it we added an *epistemic component* to the network mapping process and asked respondents questions regarding what he /she thinks every other member thinks about the structure of the network (question 1) and what he or she thinks every other member thinks he or she thinks about the structure of the network. The results (Figure 3D) indicate different members of the network have different first, second and third order beliefs about the structure of the network: what Abe thinks the network of friendship, interaction and collaboration ties 'is' can be different from what Abe thinks Cam thinks it is; both may differ from what Abe thinks Cam thinks Abe thinks it is.

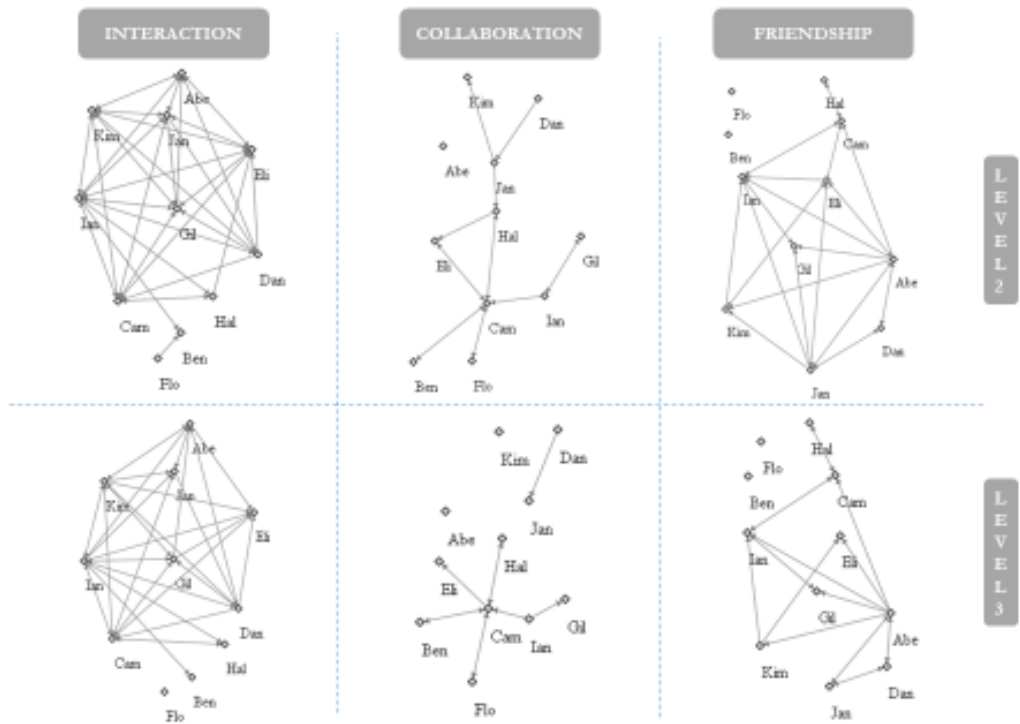

*Figure 3D. Faculty Member Interaction, Collaboration, and Friendship Networks (Levels 2&3).*

  Networks co-mobilize and coordinate around *relevant facts* or propositions describing states of affairs. We asked respondents (*N=11*) about other respondents' knowledge and b) about other respondents' knowledge about their own knowledge about a set of issues of *common and immediate concern to all faculty* regarding a proposed re-design of the area's doctoral program. We asked them to give their belief regarding the validity of propositions about critical events (delays in the opening of a new building for the school that had been scheduled for opening within the next two years, and changes to the doctoral program administered and run by area), their beliefs regarding other agents' beliefs regarding the truth value of these propositions, and their beliefs regarding other agents' beliefs regarding their own beliefs about the these propositions. The questions had *unambiguous* and *uncontroversial* 'True/False' answers (corresponding to the truth values of the propositions in question), that could be independently verified by reference to an existing document: they were statements of fact with immediate bearing upon the activities of most of the faculty members within the area. The epistemic networks generated by the answers to these questions are displayed in Figures 4A-4M.

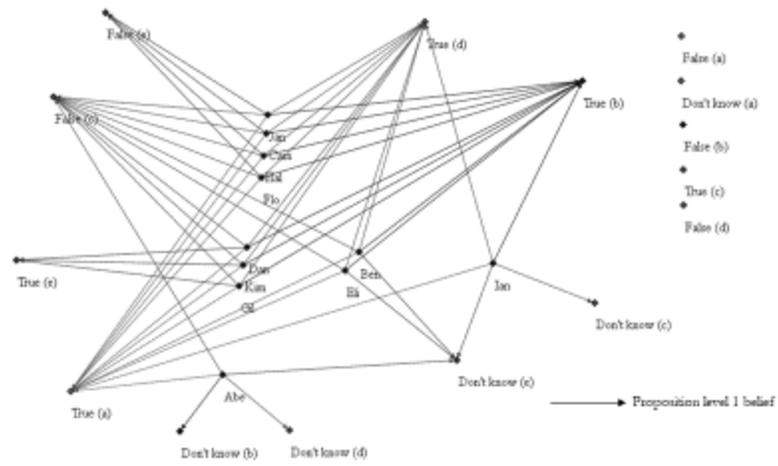

Figure 4A. Level 1 Epinet: PhD Program Propositions (a)-(e).

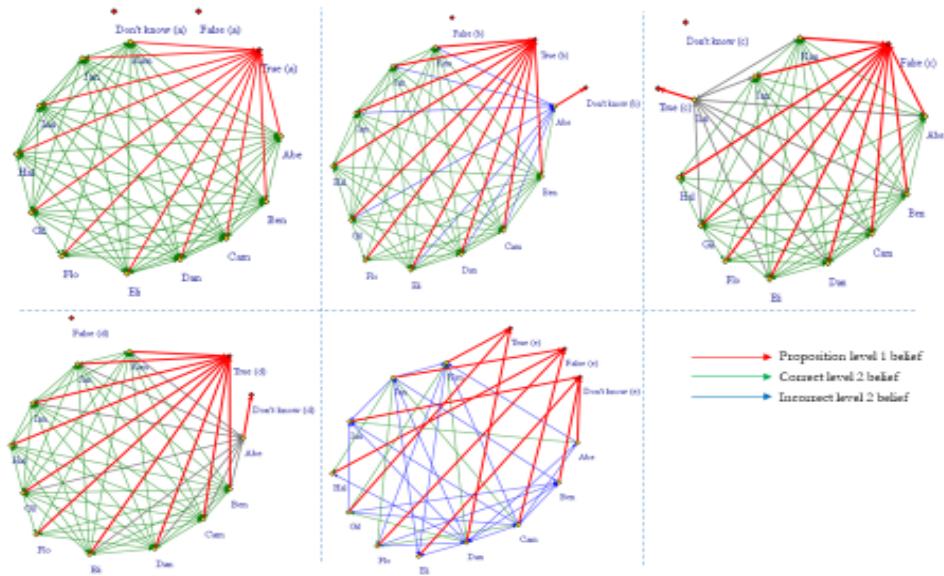

Figure 4B. Level 2 Epinets: Propositions (a)-(e).

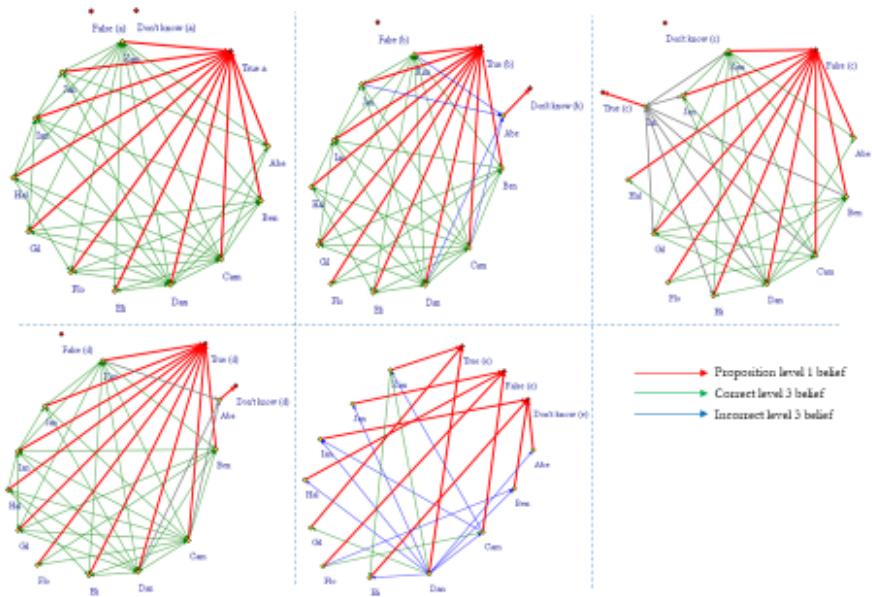

*Figure 4C. Level 3 Epinets: Propositions (a)-(e).*

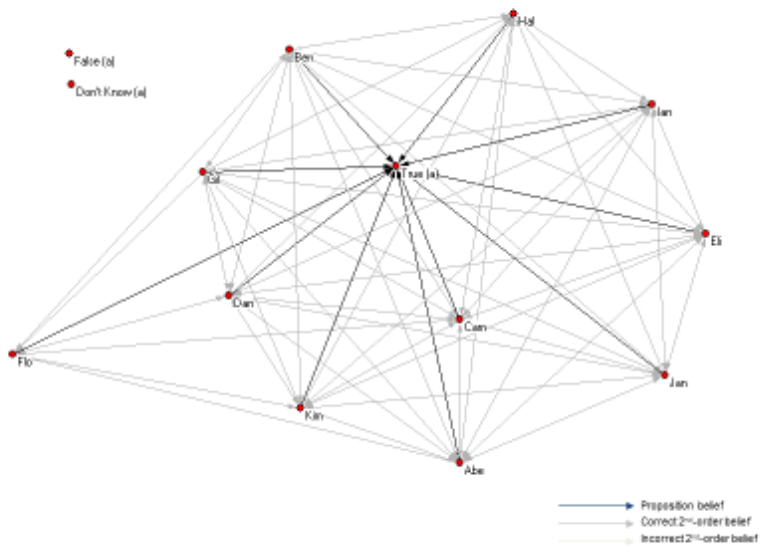

*Figure 4D. Second-order Epistemic Network. Proposition a.*

**Second-order Epistemic Network**
**Proposition b:** The proposed curriculum includes the graduate-level microeconomics sequence of courses required of graduate students in the Department of X at the University of Y: T or F?

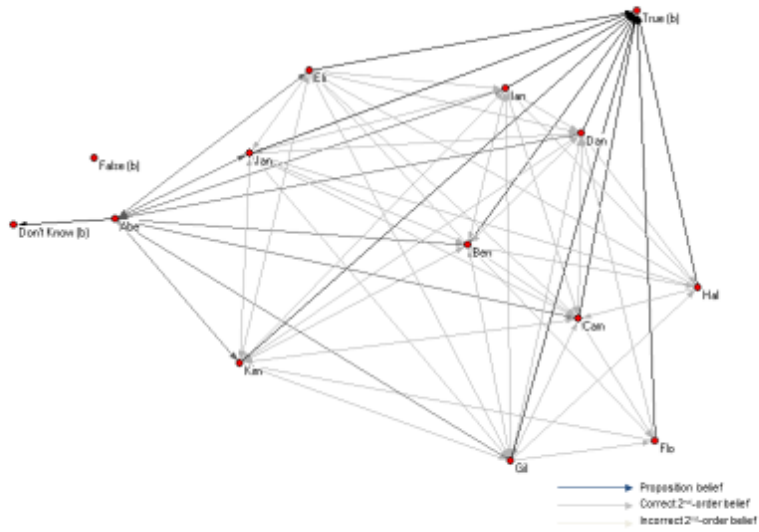

*Figure 4E. Second-order Epistemic Network. Proposition b.*

**Second-order Epistemic Network**
**Proposition c:** The proposed course curriculum requires no course work in sociology: T or F?

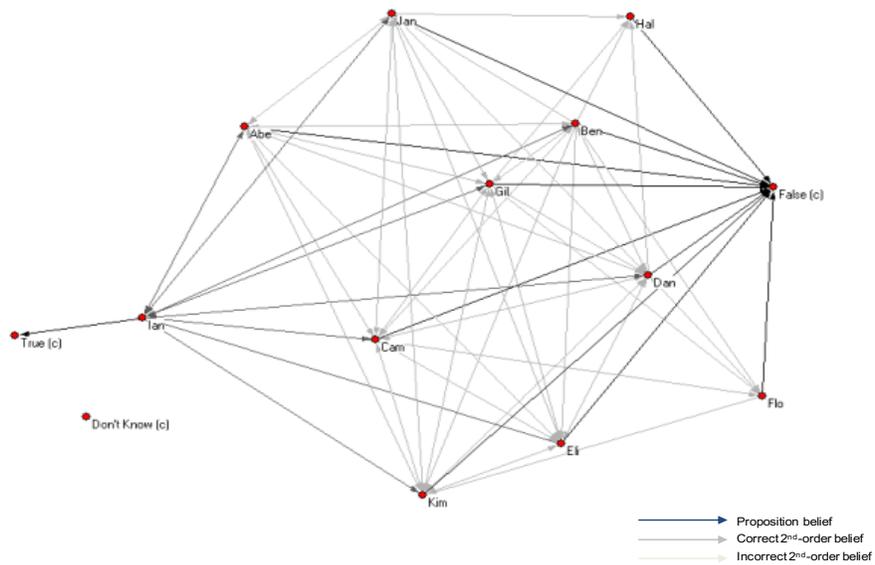

*Figure 4F. Second-order Epistemic Network. Proposition c.*

**Second-order Epistemic Network**
**Proposition d:** The proposed course curriculum includes an in-depth econometrics training sequence in addition to area-specific courses on research methods: T or F?

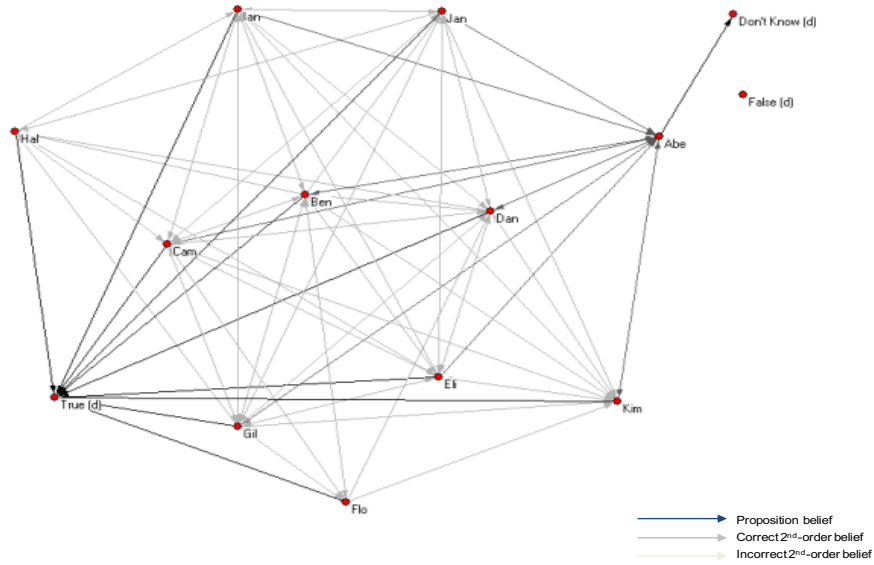

*Figure 4G. Second-order Epistemic Network. Proposition d.*

**Second-order Epistemic Network**
**Proposition e:** The proposal was discussed by the PhD Program Committee and given a conditional endorsement: T or F?

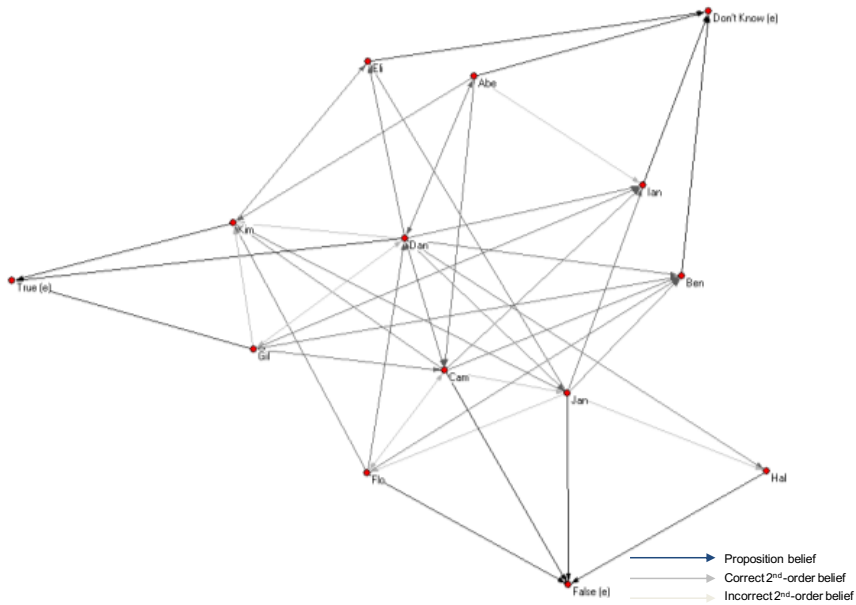

*Figure 4H. Second-order Epistemic Order. Proposition e.*

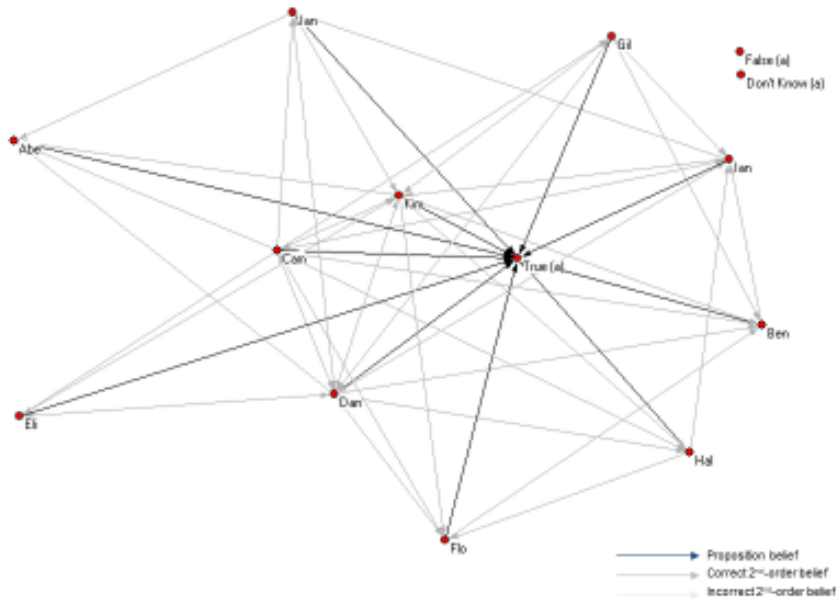

*Figure 4I. Third-order Epistemic Network. Proposition a.*

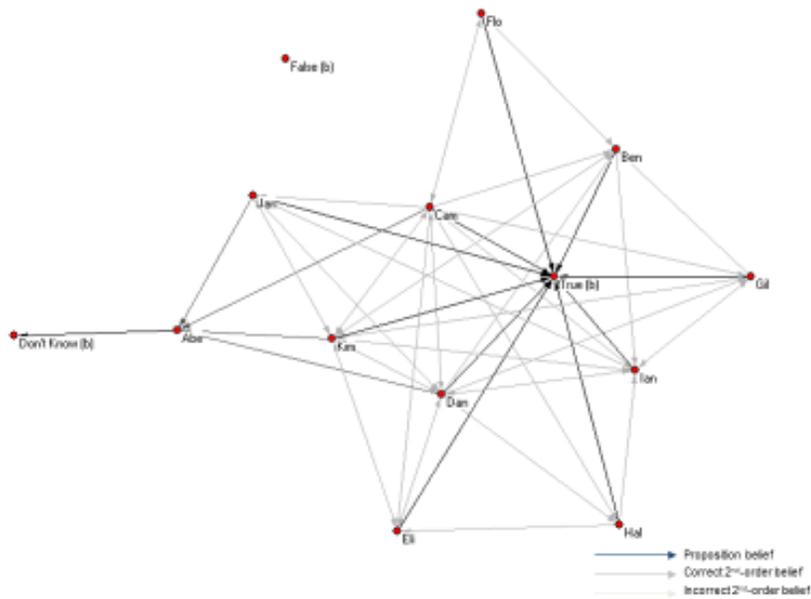

*Figure 4J. Third-order Epistemic Network. Proposition b.*

**Third-order Epistemic Network**
**Proposition c:** The proposed course curriculum requires no course work in sociology: T or F?

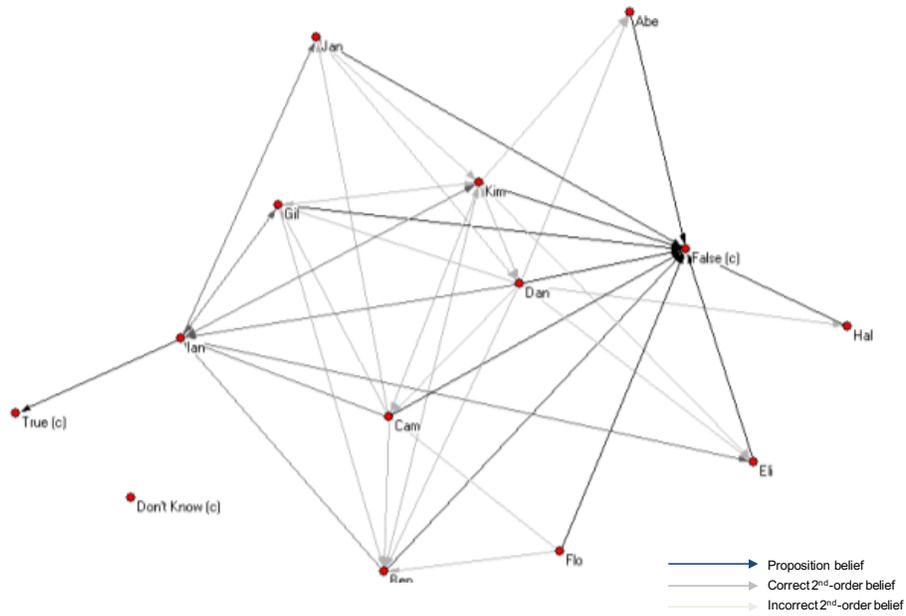

*Figure 4K. Third-order Epistemic Network. Proposition c.*

**Third-order Epistemic Network**
**Proposition d:** The proposed course curriculum includes an in-depth econometrics training sequence in addition to area-specific courses on research methods: T or F?

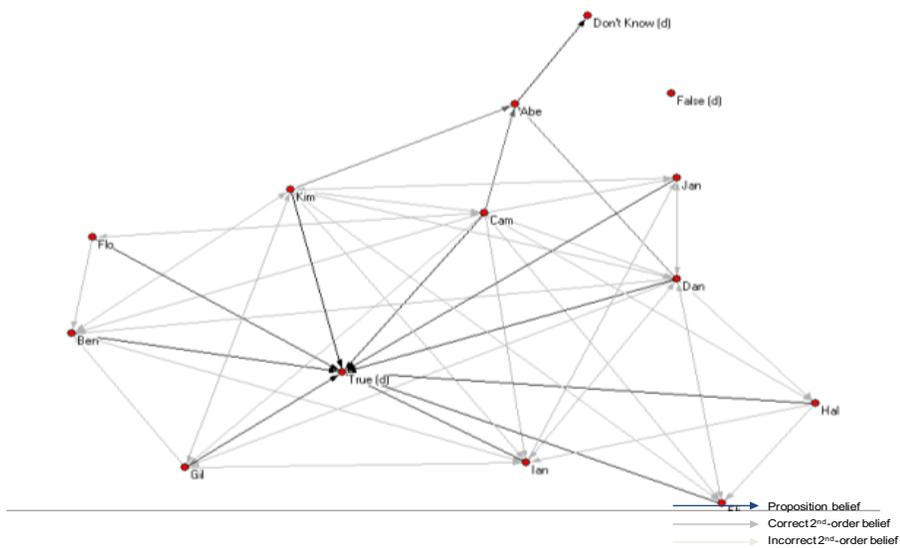

*Figure 4L. Third-order Epistemic Network. Proposition d.*

**Third-order Epistemic Network**
**Proposition e:** The proposal was discussed by the PhD Program Committee and given a conditional endorsement: T or F?

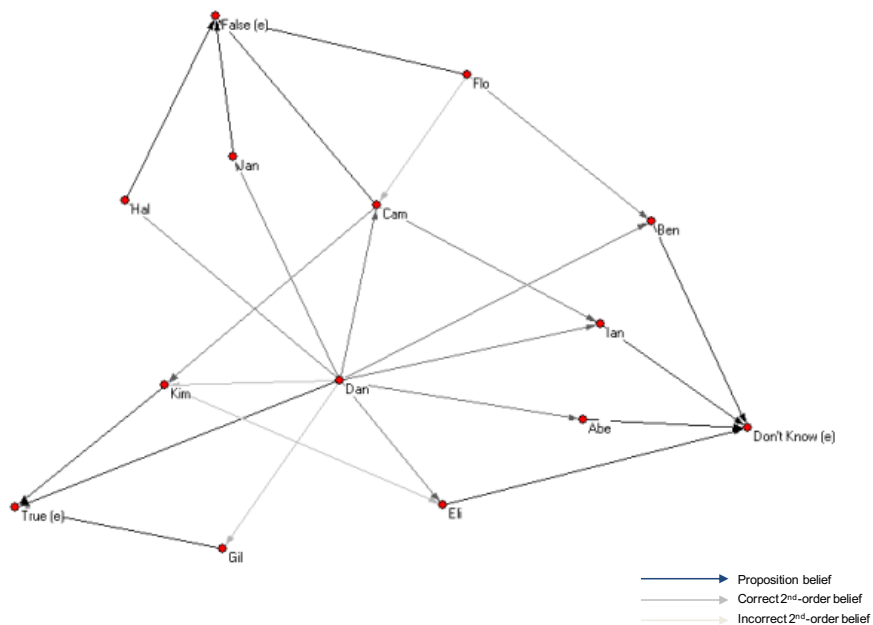

*Figure 4M. Third-order Epistemic Network. Proposition e.*

There are discrepancies among epistemic networks at different epistemic levels. If we take coherent first and second order beliefs to proxy for epistemic conditions for successful mobilization of the network around an issue and coherent first, second and third order beliefs to be a proxy for the instantiation of epistemic pre-conditions for the successful coordination of the network around focal points - the events to which the propositions refer - we see that even in a densely connected network (of friendship, collaboration and interaction-based ties) comprising agents expected to be sophisticated in terms of both network closure effects (several are social networks researchers) and the epistemic pre-conditions for convergence to the equilibria in coordination games (applied economists skilled in game theory) the low level of epistemic cohesion does not bode well for the mobilizability and coordinatability of the network.

To what extent does centrality confer advantages of a) accurate mutual knowledge (accurate knowledge about other agents' knowledge, or, level-2 knowledge), and b) accurate third-level almost-common knowledge (accurate knowledge about other agents' knowledge about the central agent's own knowledge, or, level-3 knowledge)? This question tests assumptions about the role of central agents in mobilizing and coordinating networks. The relevant results are shown in Figure 5 for level-2 and level-3 regimes, '3a-3e' refer to the five separate questions asked in the survey.

Correlations of Network Centrality and Level 2 and Level 3 Knowledge, Propositions (a)-(e)

|  |  | Level 2 knowledge | | | | | | Level 3 knowledge | | | | | |
| --- | --- | --- | --- | --- | --- | --- | --- | --- | --- | --- | --- | --- | --- |
|  |  | a | b | c | d | e | Overall | a | b | c | d | e | Overall |
| Collaboration Network | Betweenness | na | .21 | -.13 | .21 | .40 | .17 | -.14 | .06 | -.25 | .05 | .21 | -.01 |
|  | Degree | na | .38 | -.09 | .38 | .44 | .28 | -.22 | .18 | -.21 | .17 | .12 | .01 |
|  | Eigenvector | na | .45 | -.03 | .45 | .54 | .35 | -.41 | .12 | -.30 | .12 | -.10 | -.12 |
| Friendship Network | Betweenness | na | -.32 | -.41 | -.32 | .51 | -.13 | .18 | -.17 | -.08 | -.18 | .11 | -.03 |
|  | Degree | na | -.47 | -.28 | -.47 | .14 | -.27 | .34 | -.22 | .01 | -.26 | .16 | .01 |
|  | Eigenvector | na | -.40 | -.34 | -.40 | -.02 | -.29 | .41 | -.12 | .03 | -.15 | .10 | .05 |
| Interaction Network | Betweenness | na | .20 | -.83 | .20 | .37 | -.02 | -.08 | .14 | -.40 | .14 | -.50 | -.14 |
|  | Degree | na | .15 | -.63 | .15 | -.07 | -.10 | .23 | .22 | -.22 | .21 | -.20 | .05 |
|  | Eigenvector | na | -.07 | -.27 | -.07 | -.35 | -.19 | .54 | .25 | .17 | .23 | .23 | .29 |

*Figure 5. Correlations of Network Centrality and Level 2 and Level 3 Knowledge. Propositions (a)-(e).*

Overall correlations between centrality and accuracy of higher-level beliefs are weak for all three networks. All three centrality measures in the collaboration network exhibit high correlations with the central agent's accuracy of level-2 beliefs (these agents know what others know), but correlation of the same centrality measures in the collaboration network with the central agent's level-3 beliefs are weak (these agents are not likely to know what other agents know they know). There is no significant difference in the overall predictive power of different centrality measures to accuracy of level-2 and level-3 beliefs across the three networks. Agent-level centrality is not a good proxy for the degree of 'in-the-know' of the central agent, if we define 'in-the-know' as a 2-level or 3-level interactive knowledge structure.

      To what degree are the cliques of the interaction, friendship and collaboration networks also 'social pockets' of accurate (a) level-2 and (b) level-3 knowledge? We correlated the accuracy of intra-clique level-2 and level-3 knowledge with the accuracy of level-2 and level-3 knowledge for the entire network, for each clique. We analyzed all of the cliques of the collaboration, friendship and interaction networks (Figure 6) to understand the degree of coherence of level-2 and level-3 beliefs within each clique and measured epistemic coherence of level-2 and level-3 beliefs as the percentage of times an agent's level-1,2,3 knowledge matched the level-1,2,3 knowledge of another agent within the clique: A hundred per cent match represents a state of the world in which the two agents agree on the answer to the question, each knows this fact, and each knows that the other knows this fact.

## Clique versus Network Belief Coherence, Propositions (a)-(e)

| Clique Members | | | Clique Issue cohesion | | | | | | vs. Network Issue Cohesion | | | | | |
|---|---|---|---|---|---|---|---|---|---|---|---|---|---|---|
| | | | a | b | c | d | e | Overall | a | b | c | d | e | Overall |
| All Network Members | | Level 2 N<br>% | 90<br>52% | 68<br>62% | 64<br>58% | 67<br>61% | 10<br>9% | 299<br>54% | | | | | | |
| | | Level 3 N<br>% | 50<br>45% | 41<br>37% | 27<br>25% | 42<br>38% | 3<br>3% | 163<br>30% | | | | | | |
| Collaboration Network | Can Eli Hal | Level 2 N<br>% | 5<br>83% | 4<br>67% | 2<br>33% | 4<br>67% | 0<br>0% | 15<br>50% | 2% | 5% | -25% | 6% | -8% | -4% |
| | | Level 3 N<br>% | 4<br>67% | 3<br>50% | 0<br>0% | 4<br>67% | 0<br>0% | 11<br>37% | 21% | 13% | -25% | 28% | -3% | 7% |
| Friendship Network | Abe Eli Ian Jan Kim | Level 2 N<br>% | 19<br>95% | 11<br>55% | 12<br>60% | 11<br>55% | 1<br>5% | 54<br>54% | 13% | -7% | 2% | -6% | -4% | 0% |
| | | Level 3 N<br>% | 9<br>45% | 7<br>35% | 5<br>25% | 7<br>35% | 0<br>0% | 28<br>28% | 0% | -2% | 0% | -3% | -3% | -2% |
| | Abe Can Eli Ian | Level 2 N<br>% | 11<br>92% | 5<br>42% | 6<br>50% | 5<br>42% | 1<br>8% | 28<br>47% | 10% | -20% | -8% | -19% | -1% | -8% |
| | | Level 3 N<br>% | 3<br>25% | 2<br>17% | 0<br>0% | 2<br>17% | 0<br>0% | 7<br>12% | -20% | -21% | -25% | -22% | -3% | -18% |
| | Abe Gil Ian Jan | Level 2 N<br>% | 12<br>100% | 6<br>50% | 6<br>50% | 6<br>50% | 1<br>8% | 31<br>52% | 18% | -12% | -8% | -11% | -1% | -3% |
| | | Level 3 N<br>% | 4<br>33% | 3<br>25% | 0<br>0% | 3<br>25% | 0<br>0% | 10<br>17% | -12% | -12% | -25% | -13% | -3% | -13% |
| | Abe Dan Jan | Level 2 N<br>% | 6<br>100% | 2<br>33% | 6<br>100% | 2<br>33% | 0<br>0% | 16<br>53% | 18% | -25% | 42% | -25% | -9% | -1% |
| | | Level 3 N<br>% | 4<br>67% | 2<br>33% | 3<br>50% | 2<br>33% | 0<br>0% | 11<br>37% | 21% | -4% | 25% | -5% | -3% | 7% |
| Interaction Network | Can Dan Eli Gil Ian Jan Kim | Level 2 N<br>% | 41<br>98% | 41<br>98% | 30<br>71% | 41<br>98% | 5<br>12% | 158<br>75% | 16% | 36% | 13% | 37% | 3% | 21% |
| | | Level 3 N<br>% | 29<br>69% | 28<br>67% | 18<br>43% | 28<br>67% | 2<br>5% | 105<br>50% | 24% | 29% | 18% | 28% | 2% | 20% |
| | Can Eli Hal Ian | Level 2 N<br>% | 3<br>75% | 8<br>67% | 2<br>17% | 8<br>67% | 0<br>0% | 21<br>45% | -7% | 5% | -42% | 6% | -8% | -9% |
| | | Level 3 N<br>% | 6<br>50% | 5<br>42% | 0<br>0% | 6<br>50% | 0<br>0% | 17<br>28% | 5% | 4% | -25% | 12% | -3% | -1% |
| | Abe Dan Eli Gil Ian Jan Kim | Level 2 N<br>% | 41<br>98% | 29<br>69% | 30<br>71% | 29<br>69% | 5<br>12% | 134<br>64% | 16% | 7% | 13% | 8% | 3% | 9% |
| | | Level 3 N<br>% | 24<br>57% | 20<br>48% | 15<br>36% | 20<br>48% | 2<br>5% | 81<br>39% | 12% | 10% | 11% | 9% | 2% | 9% |
| | Ben Eli Ian | Level 2 N<br>% | 4<br>67% | 4<br>67% | 1<br>17% | 4<br>67% | 0<br>0% | 13<br>43% | -15% | 5% | -42% | 6% | -8% | -11% |
| | | Level 3 N<br>% | 1<br>17% | 1<br>17% | 0<br>0% | 1<br>17% | 0<br>0% | 3<br>10% | -20% | -21% | -25% | -22% | -3% | -20% |
| | Dan Gil Kim | Level 2 N<br>% | 6<br>100% | 6<br>100% | 6<br>100% | 6<br>100% | 3<br>53% | 29<br>97% | 18% | 38% | 42% | 38% | 74% | 42% |
| | | Level 3 N<br>% | 6<br>100% | 6<br>100% | 6<br>100% | 6<br>100% | 2<br>33% | 26<br>87% | 55% | 63% | 75% | 62% | 31% | 57% |

*Figure 6. Clique versus Network Belief Coherence. Propositions (a)-(e).*

We computed overall agreement within the entire network and the difference between intra-clique and network levels of agreement and the number of matches for the entire network that would result from a binomial distribution of epistemic states within the network. Differences between intra-clique agreement and overall agreement within the network are low, with a few exceptions, the most notable of which is the (Dan, Gil, Kim) clique of the interaction network. This clique is remarkable. It is not made up of central agents. Moreover, although the coherence among the agents in the clique is high, *their beliefs are incorrect*; had the answers to the questions posed to the agents within the clique functioned as a set of focal points in a coordination game they would have coordinated on the *wrong signal*.

     These findings suggest network-topological ('cliques' in interaction networks) and structural ('actor centrality') features *by themselves* should not be interpreted as good predictors of network cohesion or actor influence. Neither clique membership nor agent-level centrality in themselves confer the informational advantages that should accrue to the central and the tightly connected that would allow them to 'work the network'. If this informational advantage is critical to the phenomena of brokerage and closure that constitute social capital, then a map of the epistemic conditions of social capital formation and maintenance is required.

For instance: Central agents 'know and are known (to be central) by' others in their network, but in order for their position to enable them to *mobilize* and *coordinate* a network, an epistemic description thereof is useful.

In an epinet:

- A *mobilizer* will have insight into (a) the structure of the network, (b) others' perceptions of the structure of the network, and (c) the set of mobilization rules and thresholds that are 'mutual knowledge' among various sub-networks within the network. She will understand how to get sparsely connected groups to act cohesively by making the 'right introductions' – which will decrease the barriers to mobilization (the difference between the minimum number of others that must mobilize in order for an agent to mobilize and the actual number of others whom the agent knows will mobilize if he does) for the individual agents within the network most likely to co-mobilize given mutual knowledge of intent;
- A *coordinator* will know enough about what other agents know and what they know that others know to predict which of several possible focal points in a coordination game is most salient.

Using epinet maps, we can ask: *Does* centrality confer upon the central the kind of epistemic advantages that are sufficient for mobilizing and coordinating their networks? - rather than often incorrectly *assuming* it does.

Supplemental Information

I. Questionnaire.

**1.** For each XXX Area faculty member listed in the table below, place an '**X**' in the column(s) corresponding to the number(s) of the statement(s) you believe to be true of your relationship with him/her:

1. I am collaborating with the faculty member on coauthored academic work;
2. I interact with the faculty member about once a week;
3. I interact with the faculty member about once a day;
4. I consider the faculty member to be a personal friend.

| Faculty Member | | 1 | 2 | 3 | 4 | A |
|---|---|---|---|---|---|---|
| Abe | Abe | | | | | |
| Ben | Ben | | | | | |
| Cam | Cam | | | | | |
| Dan | Dan | | | | | |
| Eli | Eli | | | | | |
| Flo | Flo | | | | | |
| Gil | Gil | | | | | |
| Hal | Hal | | | | | |
| Ian | Ian | | | | | |
| Jan | Jan | | | | | |
| Kim | Kim | | | | | |

Now, for **all faculty members** listed above: 1) place a '**1**' in column A beside the name of each faculty member who you believe would characterize your relationship the same way that you did and 2) place a '**2**' in column A beside the name of each faculty member who you believe knows how you characterized your relationship with them.

Note: If you believe both 1) and 2) to be true for a particular faculty member, place **both** a '**1**' and a '**2**' in column A beside their name.

---

2. At the end of each of the following statements, please indicate the name of the Strategy Area faculty member from the list below whom you believe to hold the network position indicated.

a. The faculty member with the highest degree centrality in the Strategy Area friendship network (i.e., has the most friendship ties with others) is: _______________________

b.  The faculty member with the highest betweenness centrality in the XXX Area interaction network (i.e., has the largest fraction of shortest paths between faculty members pass through him/her) is: _______________________

c.  The faculty member with the highest Bonacich centrality in the XXX Area academic collaboration network (i.e., has the *most* ties to *the most central* faculty in the collaboration network) is: _______________________

Now, for **all faculty members** listed below: 1) place a **'1'** in the column beside the name of each faculty members who you believe would give the same responses as you did to statements 2a–2c and 2) place a **'2'** in the column beside the name of each faculty member who you believe knows what your responses are to statements 2a–2c.

Note: If you believe both 1) and 2) to be true for a particular faculty member, place **both** a '1' and a '2' in the appropriate column beside their name.

| Faculty Member | 2a | 2b | 2c |
| --- | --- | --- | --- |
| Abe | | | |
| Ben | | | |
| Cam | | | |
| Dan | | | |
| Eli | | | |
| Flo | | | |
| Gil | | | |
| Hal | | | |
| Ian | | | |
| Jan | | | |
| Kim | | | |

3. For each of these following statements, please indicate whether or not you know it to be true by placing a 'T' or false 'F' following the statement.

a.  A new proposed curriculum for the PhD in XXX was articulated by ____ and _______ and circulated last month to Area members: **T or F?**

b.  The proposed curriculum includes the graduate-level microeconomics sequence of courses required of graduate students in the Department of XXX at the University of XXX: **T or F?**

c.  The proposed course curriculum requires no course work in sociology: **T or F?**

d.  The proposed course curriculum includes an in-depth econometrics training sequence in addition to area-specific courses on research methods: **T or F?**

e.  The proposal was discussed by the PhD Program Committee and given a conditional endorsement: **T or F?**

f.  Plans for the new Rotman School building have been put on hold pending resolution of regulatory issues: **T or F?**

Now, for **all faculty members** listed below: 1) place a **'1'** in the column beside the name of each faculty member who you believe would give the same responses as you did to statements 3a–3f and 2) place a **'2'** in the column beside the name of each faculty member who you believe knows what your responses are to statements 3a–3f.

Note: If you believe both 1) and 2) to be true for a particular faculty member, place **both** a '1' and a '2' in the appropriate column beside their name.

| Faculty Member | 3a | 3b | 3c | 3d | 3e | 3f |
|---|---|---|---|---|---|---|
| Abe | | | | | | |
| Ben | | | | | | |
| Cam | | | | | | |
| Dan | | | | | | |
| Eli | | | | | | |
| Flo | | | | | | |
| Gil | | | | | | |
| Hal | | | | | | |
| Ian | | | | | | |
| Jan | | | | | | |
| Kim | | | | | | |